\documentclass[letterpaper, 10 pt, conference]{ieeeconf}%
\usepackage{bm}
\usepackage{amssymb}
\usepackage{epsfig}
\usepackage{subfigure}
\usepackage{amsmath}
\usepackage{xcolor}
\usepackage{graphicx}
\usepackage{epstopdf}
\usepackage{amsfonts}
\usepackage{indentfirst}%
\usepackage[ruled]{algorithm2e}
\usepackage{microtype}
\usepackage{booktabs} 
\usepackage{array}

\newif\ifmarginnotes
\marginnotestrue 
\ifmarginnotes
    \newcommand{\RAnote}[1]{%
    \marginpar{\tiny\color{red}#1}}
\else
    \newcommand{\RAnote}[1]{}
\fi

\providecommand{\U}[1]{\protect\rule{.1in}{.1in}}
\newtheorem{problem}{\textbf{Problem}}
\newtheorem{definition}{\textbf{Definition}}

\newtheorem{theorem}{\rm\textbf{Theorem}}

\newtheorem{remark}{\rm\textbf{Remark}}
\IEEEoverridecommandlockouts
\begin{document}

\title{{\LARGE \textbf{Safe Neural Control for Non-Affine Control Systems with \\Differentiable Control Barrier Functions}}}
\author{Wei Xiao, Ross Allen, and Daniela Rus\thanks{This work was partially supported by Capgemini Engineering. This research was also sponsored by the United States Air Force Research Laboratory and the United States Air Force Artificial Intelligence Accelerator and was accomplished under Cooperative Agreement Number FA8750-19-2-1000. The views and conclusions contained in this document are those of the authors and should not be interpreted as representing the official policies, either expressed or implied, of the United States Air Force or the U.S. Government. The U.S. Government is authorized to reproduce and distribute reprints for Government purposes, notwithstanding any copyright notation herein. This research was also supported in part by the AI2050 program at Schmidt Futures (Grant G-965 22-63172). }
\thanks{Wei Xiao and Daniela Rus are with CSAIL, MIT, Ross Allen is with the MIT Lincoln Lab. \texttt{{\small weixy@mit.edu, ross.allen@ll.mit.edu, rus@csail.mit.edu}}.}}
\maketitle

\begin{abstract}
This paper addresses the problem of safety-critical control for non-affine control systems. It has been shown that optimizing quadratic costs subject to state and control
constraints can be sub-optimally reduced to a sequence of quadratic programs (QPs) by using
Control Barrier Functions (CBFs). Our recently proposed High Order CBFs (HOCBFs) can accommodate constraints of arbitrary relative degree. The main challenges in this approach are that it requires affine control dynamics and the solution of the CBF-based QP is sub-optimal since it is solved point-wise. 
To address these challenges, we incorporate higher-order CBFs into neural ordinary differential equation-based learning models as differentiable CBFs to guarantee safety for non-affine control systems. The differentiable CBFs are trainable in terms of their parameters, and thus, they can address the conservativeness of CBFs such that the system state will not stay unnecessarily far away from safe set boundaries. Moreover, the imitation learning model is capable of learning complex and optimal control policies that are usually intractable online. 
 We illustrate the effectiveness of the proposed framework on  LiDAR-based autonomous driving and compare it with existing methods.
\end{abstract}

\thispagestyle{empty} \pagestyle{empty}



\section{INTRODUCTION}

\label{sec:intro}

Optimal control problems with safety requirements are central to increasingly widespread safety critical autonomous and cyber physical systems. Control barrier functions enforcing safety have received increased attention in recent years \cite{Aaron2014} \cite{Glotfelter2017} \cite{Xiao2019} due to their high computational efficiency in dealing with {affine-control nonlinear systems.}

Barrier functions (BFs) are Lyapunov-like functions \cite{Tee2009},
\cite{Wieland2007}, whose use can be traced back to optimization problems
\cite{Boyd2004}. More recently, they have been employed to prove set
invariance \cite{Aubin2009}, \cite{Prajna2007}, \cite{Wisniewski2013} and for
multi-objective control \cite{Panagou2013}. %
Tee et al.~\cite{Tee2009} proved
that if a BF for a given set satisfies Lyapunov-like conditions, then the set
is forward invariant. A less restrictive form of a BF, which is allowed to
decrease when far away from the boundary of the set, was proposed by Ames et al.~\cite{Aaron2014}. 
Control BFs (CBFs) are extensions of BFs for control systems, and are used to
map a constraint defined over system states to a constraint on the control
input. %
The CBFs proposed by Ames et al.~\cite{Aaron2014} and Glotfelter et al.~\cite{Glotfelter2017} work for constraints
that have relative degree one with respect to the system dynamics.  Exponential CBFs \cite{Nguyen2016} 
for arbitrarily high relative degree constraints employ input-output
linearization and find a pole placement controller with negative poles. The high order
CBF (HOCBF) proposed by Xiao and Belta~\cite{Xiao2019} is simpler (to define) and more general than the
exponential CBF \cite{Nguyen2016}.

Most works using CBFs to enforce safety are based on the assumption that the control system is affine in controls and the cost is quadratic in controls.  The time domain is discretized, and the state is assumed to be constant over each time interval. 
The optimal control problem is sub-optimally reduced to a quadratic program (QP) in each time interval and the control is kept constant for the whole interval. Using this approach, the original optimal control problem is reduced to a (possibly large) sequence of QPs - one for each interval.
There are two main challenges in the aforementioned CBF-based QP formulation: (i) the dynamics must be affine in control. Otherwise, the CBF-based optimization would be a sequence of nonlinear programs (NLPs) that are inefficient and hard to solve \cite{son2019safety} \cite{seo2022safety}. (ii) the solution of the CBF-based QP is sub-optimal since the problem is solved point-wise.

In order to reformulate constrained optimal control problems as CBF-based QPs for non-affine control systems, one can augment the system with auxiliary dynamics such that the augmented dynamics would be affine in controls \cite{slotine1991applied} \cite{XiaoCBF2022}. This is achieved at the cost of higher-relative degree CBFs, and the eventual CBFs become integral CBFs \cite{Ames2020} since the integral solution of the CBF-based QP is the control for the original non-affine control system. However, there is no formulaic procedure to define such auxiliary dynamics. The second challenge mentioned above can be addressed using the nonlinear model predictive control (NMPC) method \cite{Rawlings2018} or the inverse optimal method \cite{krstic2023inverse}. However, this may lead to NLPs with computationally expensive solutions. %
Although linearization is possible in NMPC to decrease the complexity, it may come at the cost of loss of safety guarantees. Another way to improve the optimality is to employ imitation learning \cite{wang2022learning} \cite{dawson2023safe}. 
This approach learns complex control policies that are hard to solve online, and maps the learned policies to system observations, such as the front-view RGB images in driving. The limitation of imitation learning is that there is no guaranteed safety.  BarrierNet \cite{xiao2021bnet} has been proposed to equip learning systems with guarantees, but it does not work for non-affine control systems. CBFs have been used in neural Ordinary Differential Equations (ODEs)-based learning models \cite{chen2018neural} to equip them with guarantees \cite{xiao2022forward} \cite{huang2022fi}. However, these CBFs are either used to modify the model trainable parameters such that CBFs are to be considered during training 
\cite{xiao2022forward} or not trainable in the model \cite{huang2022fi}.

To address the problem of safety-critical control for non-affine control systems and improve the optimality of the solution,  this paper contributes a continuous optimization learning method that employs imitation learning with guarantees. The proposed learning model is based on neural ODEs that excel in learning control policies~\cite{chen2018neural}. We define CBFs enforcing safety for non-affine control systems to higher-order CBFs such that the eventual CBF constraints would be linear in decision variables. Then, we incorporate these higher-order CBFs into the neural ODEs as differentiable CBFs (in terms of the CBF parameters) that are trainable to address the conservativeness of CBFs. Finally, we show that this method works efficiently for non-affine control systems with safety guarantees. 
 We illustrate our approach and compare with other methods on a LiDAR-based  autonomous driving problem.


\section{PRELIMINARIES}

\label{sec:pre}

\begin{definition}
	\label{def:classk} (\textit{Class $\mathcal{K}$ function} \cite{Khalil2002}) A
	continuous function $\alpha:[0,a)\rightarrow[0,\infty), a > 0$ is said to
	belong to class $\mathcal{K}$ if it is strictly increasing and $\alpha(0)=0$. A continuous function $\beta:\mathbb{R}\rightarrow\mathbb{R}$ is said to belong to extended class $\mathcal{K}$ if it is strictly increasing and $\beta(0)=0$.
\end{definition}

Consider an affine control system (assumed to be affine in control only in this section) of the form:
\begin{equation}
\dot{\bm{x}}=f(\bm x)+g(\bm x)\bm u \label{eqn:affine}%
\end{equation}
where $\bm x\in X\subset\mathbb{R}^{n}$, $f:\mathbb{R}^{n}\rightarrow\mathbb{R}^{n}$
and $g:\mathbb{R}^{n}\rightarrow\mathbb{R}^{n\times q}$ are
Lipschitz continuous, and $\bm u\in U\subset\mathbb{R}^{q}$ is the control constraint set
defined as ($\bm u_{min},\bm u_{max}\in\mathbb{R}^{q}$):
\begin{equation}
U:=\{\bm u\in\mathbb{R}^{q}:\bm u_{min}\leq\bm u\leq\bm u_{max}\},
\label{eqn:control}%
\end{equation}
where the inequalities are interpreted element-wise.

\subsection{Control Barrier Functions and BarrierNet}

We first introduce the concept of control barrier functions that are control synthesis tools for safe autonomous systems, and then briefly introduce BarrierNet that enables end-to-end safe learning.

\begin{definition}
	\label{def:forwardinv} A set $C\subset\mathbb{R}^{n}$ is forward invariant for
	system (\ref{eqn:affine}) if its solutions {for some $\bm u\in U$} starting at any $\bm x(0) \in C$
	satisfy $\bm x(t)\in C,$ $\forall t\geq0$.
\end{definition}

\begin{definition}
	\label{def:relative} (\textit{Relative degree} \cite{Khalil2002}) The relative degree of a
	(sufficiently many times) differentiable function $b:\mathbb{R}^{n}%
	\rightarrow\mathbb{R}$ with respect to system (\ref{eqn:affine}) is the number
	of times it needs to be differentiated along its dynamics until any component of the control
	$\bm u$ explicitly shows  in the corresponding derivative.
\end{definition}

In this paper, we assume that if there exists $\bm x$ such that the control shows up in the derivative of $b$, then it shows up for all $\bm x$. 
Since function $b$ is used to define a constraint $b(\bm
x)\geq0$, we will also refer to the relative degree of $b$ as the relative
degree of the constraint.
For a constraint $b(\bm x)\geq0$ with relative
degree $m$, $b:\mathbb{R}^{n}\rightarrow\mathbb{R}$, and $\psi_{0}(\bm
x):=b(\bm x)$, we define a sequence of functions $\psi_{i}:\mathbb{R}%
^{n}\rightarrow\mathbb{R},i\in\{1,\dots,m\}$:
\begin{equation}
\begin{aligned} \psi_i(\bm x) := \dot \psi_{i-1}(\bm x) + \alpha_i(\psi_{i-1}(\bm x)),i\in\{1,\dots,m\}, \end{aligned} \label{eqn:functions}%
\end{equation}
where $\alpha_{i}(\cdot),i\in\{1,\dots,m\}$ denotes a $(m-i)^{th}$ order
differentiable class $\mathcal{K}$ function.

We further define a sequence of sets $C_{i}, i\in\{1,\dots,m\}$ associated
with (\ref{eqn:functions}) in the form:
\begin{equation}
\label{eqn:sets}\begin{aligned} C_i := \{\bm x \in \mathbb{R}^n: \psi_{i-1}(\bm x) \geq 0\}, i\in\{1,\dots,m\}. \end{aligned}
\end{equation}

\begin{definition}
\label{def:hocbf} (\textit{High Order Control Barrier Function (HOCBF)}
\cite{Xiao2019}) Let $C_{1}, \dots, C_{m}$ be defined by (\ref{eqn:sets}%
) and $\psi_{1}(\bm x), \dots, \psi_{m}(\bm x)$ be defined by
(\ref{eqn:functions}). A function $b: \mathbb{R}^{n}\rightarrow\mathbb{R}$ is
a High Order Control Barrier Function (HOCBF) of relative degree $m$ for
system (\ref{eqn:affine}) if there exist $(m-i)^{th}$ order differentiable
class $\mathcal{K}$ functions $\alpha_{i},i\in\{1,\dots,m-1\}$ and a class
$\mathcal{K}$ function $\alpha_{m}$ such that 
{\small\begin{equation}
\label{eqn:constraint}\begin{aligned} \sup_{\bm u\in U}[L_f^{m}b(\bm x) + L_gL_f^{m-1}b(\bm x)\bm u + O(b(\bm x)) + \alpha_m(\psi_{m-1}(\bm x))] \geq 0, \end{aligned}
\end{equation}
}for all $\bm x\in C_{1} \cap,\dots, \cap C_{m}$. In
(\ref{eqn:constraint}), the left part is actually $\psi_m(\bm x)$, $L_{f}^{m}$ ($L_{g}$) denotes Lie derivatives along
$f$ ($g$) $m$ (one) times, and $O(b(\bm x)) = \sum_{i = 1}^{m-1}L_f^i(\alpha_{m-i}\circ\psi_{m-i-1})(\bm x).$ 
\end{definition}

The HOCBF is a general form of the relative degree one CBF \cite{Aaron2014},
\cite{Glotfelter2017}, i.e., setting $m=1$ reduces the HOCBF to
the common CBF form:
\begin{equation}\label{eqn:cbf0}
L_fb(\bm x) + L_gb(\bm x)\bm u + \alpha_1(b(\bm x))\geq 0,
\end{equation}
 and it is also a general form of the exponential CBF
\cite{Nguyen2016}.

\begin{theorem}
\label{thm:hocbf} (\cite{Xiao2019}) Given an HOCBF $b(\bm x)$ from Def.
\ref{def:hocbf} with the associated sets $C_{1}, \dots, C_{m}$ defined
by (\ref{eqn:sets}), if $\bm x(0) \in C_{1} \cap,\dots,\cap C_{m}$,
then any Lipschitz continuous controller $\bm u(t)\in U$ that satisfies the constraint in
(\ref{eqn:constraint}), $\forall t\geq0$ renders $C_{1}\cap,\dots,
\cap C_{m}$ forward invariant for system (\ref{eqn:affine}).
\end{theorem}

Many existing works \cite{Aaron2014}, \cite{Nguyen2016}
combine CBFs for systems with relative degree one with quadratic costs to form
optimization problems. Time is discretized and an optimization problem with
constraints given by the CBFs (inequalities in (\ref{eqn:constraint}%
)) is solved at each time step. Note that these
constraints are linear in control since the state value is fixed at the
beginning of the interval, therefore, each optimization problem is a quadratic
program (QP) {if the cost is quadratic in the control.} The optimal control obtained by solving each QP is applied at
the current time step and held constant for the whole interval. The state is
updated using dynamics (\ref{eqn:affine}), and the procedure is repeated. Formally, suppose we wish to minimize a cost function $u^T H \bm u$ for system (\ref{eqn:affine}), where $H\in\mathbb{R}^{q\times q}$ is positive definite,  the CBF-based QP is defined as follows. { We 
	partition a time interval $[0,T]$  into a set of equal time intervals $\{[0, \Delta t), [\Delta t,2\Delta t),\dots\}$, where $\Delta t > 0$. In each interval $[\omega \Delta t, (\omega+1) \Delta t)$ ($\omega = 0,1,2,\dots$), we assume the control is constant (i.e., the overall control will be piece-wise constant).
	Then {at $t = \omega \Delta t$,} we solve the QP:
	\begin{equation} \label{eqn:obj}
	\begin{aligned}
	\min_{\bm u(\omega \Delta t)} &\bm u^T(\omega \Delta t) H \bm u(\omega \Delta t) \\
	&\text{s.t. }\bm u_{min}\leq\bm u\leq\bm u_{max},\\
	L_f^{m}b(\bm x) + [L_g&L_f^{m-1}b(\bm x)]\bm u \!+\! O(b(\bm x)) + \alpha_m(\psi_{m-1}(\bm x)) \geq 0.
	\end{aligned}
	\end{equation}
	This method works conditioned on the
fact that the dynamics (\ref{eqn:affine}) are affine in control. Otherwise, the above optimization would be a sequence of NLPs that are hard and inefficient to solve. However, there are a lot of systems whose dynamics are not affine in control, such as the bicycle and drone models. In this paper, we show how we can efficiently guarantee safety for non-affine control systems, as well as directly infer safe control from high dimensional observations using imitation learning.

\noindent \textbf{BarrierNet.} A BarrierNet \cite{xiao2021bnet} is based on the CBF-based QP (\ref{eqn:obj}), and it incorporates the optimization as a trainable layer in the neural network. We can find the loss of the output (solution) of the optimization layer with respect to all the optimization hyper parameters using the Karush–Kuhn–Tucker conditions. Thus. all the parameters in the cost, such as $H$, and the class $\mathcal{K}$ functions in the HOCBF can be trained by the data instead of by hand-tuning. In this way, we can make the HOCBF adaptive to the observation of the system, as well as addressing the conservativeness introduced by HOCBFs. The HOCBF is differentiable in terms of its parameters in a BarrierNet, and thus, we call it a differentiable CBF. Similar to existing CBF methods, one significant limitations of the BarrierNet is that it cannot work for non-affine systems.

\subsection{Neural ODEs}
\label{sec:bkg}

A neural ordinary differential equation (ODE) \cite{chen2018neural} is defined as :
\begin{equation}\label{eqn:NN}
\dot {\bm x}(t) = f_{\vartheta}(\bm x(t)),
\end{equation}
where  $\bm x \in \mathbb{R}^n$ is the state and $\dot{\bm x}$ denotes the time derivative of $\bm x$, $n\in \mathbb{N}$ is state dimension, $f_{\vartheta}:\mathbb{R}^n \rightarrow\mathbb{R}^n$ is a neural network model parameterized by $\vartheta$. 
 The output of the neural ODE is the integral solution of (\ref{eqn:NN}). 
It can also include external input (e.g. observation vector), where the model is defined as:
\begin{equation}\label{eqn:NN_control}
\dot {\bm x}(t) = f_{\vartheta}^{'}(\bm x(t), \textbf{I}(t)),
\end{equation}
where $\textbf{I}(t) \in\mathbb{R}^{d}$, $d \in\mathbb{N}$ is external input dimension, $f_{\vartheta}^{'}:\mathbb{R}^n\times \mathbb{R}^{d} \rightarrow\mathbb{R}^n$ is a neural network model parameterized by $\vartheta$. 

The limitation of neural ODEs is that {they have no safety guarantees for control systems,}
which prevents their applications for safety-critical systems. In this work, we address this issue using the proposed learning framework.

\section{PROBLEM FORMULATION AND APPROACH}
\label{sec:prob}
We consider a non-affine control system whose dynamics are defined as:
\begin{equation} \label{eqn:system}
    \dot {\bm x} = h(\bm x, \bm u),
\end{equation}
where $h:\mathbb{R}^n\times \mathbb{R}^q \rightarrow \mathbb{R}^n$ is locally Lipschitz continuous.

$\textbf{Objective}$: (Minimizing cost) Consider an optimal control problem for system (\ref{eqn:system}) with the cost:
\begin{equation}\label{eqn:cost}
\min_{\bm u(t)}\int_{0}^{T}\mathcal{C}(||\bm u(t)||)dt + p_0||\bm x(T) - \bm K||^2
\end{equation}
where $T > 0, p_0 > 0, \bm K\in\mathbb{R}^n$, $||\cdot||$ denotes the 2-norm of a vector, $\mathcal{C}(\cdot)$ is a strictly increasing function of its argument. We may consider penalizing the errors of all the states with respect to $\bm K$ in the above to make it more general.

$\textbf{Safety requirements}$: 
System (\ref{eqn:system}) should always satisfy a safety requirement:
\begin{equation} \label{eqn:safetycons}
b(\bm x(t))\geq 0, \forall t\in[0,T],
\end{equation}
where $b: \mathbb{R}^n\rightarrow\mathbb{R}$ is continuously differentiable and has relative degree $m\in\mathbb{N}$ with respect to system (\ref{eqn:system}).

$\textbf{Control constraints}$: The control of the real system should always satisfy control bounds in the form of (\ref{eqn:control}).

\begin{problem}\label{prob:general}
	Given a real-time observation $\bm I\in\mathbb{R}^d$ ($\bm I$ could be the sensor information or the state $\bm x$, $d\in\mathbb{N}$ is the dimension of the observation) for system (\ref{eqn:system}), find an \textit{online} control policy for system (\ref{eqn:system}) such that the cost (\ref{eqn:cost}) is minimized, and constraints (\ref{eqn:safetycons}) and (\ref{eqn:control}) are satisfied.
\end{problem}

\textbf{Approach:} Our approach to solve Problem \ref{prob:general} is based on the proposed safety-guaranteed machine learning method. Specifically, we employ the Nonlinear Model Predictive Control (NMPC) method to estimate the optimal control $\bm u^*$ of Problem \ref{prob:general} \textit{offline} given {an observation $\bm I$}.
In this way, we collect labeled training data set in the form of $(\bm x, \bm I, \bm u^*)$ where $\bm u^*$ is the optimal control label corresponding to $(\bm x, \bm I)$. Then, we construct a continuous optimization learning model using neural ODEs and BarrierNet that is trained by the collected data, and that can be deployed for online control. We {provably show}
the safety guarantees of the learning-based control for the non-affine control system (\ref{eqn:system}).

\section{Safe Neural Control}
\label{sec:snc}

In this section, we show how we can solve Problem \ref{prob:general} using a machine learning-based method that can guarantee system safety. We start with a motivation example showing why existing CBF methods may fail to work for system (\ref{eqn:system}).

\subsection{Motivating Example}
\label{sec:exam}

Consider a bicycle model defined as:
\begin{equation} \label{eqn:bicycle}
    \begin{aligned}
        \dot x &= v\cos\theta,\qquad \dot y = v\sin\theta,\\
        \dot \theta &= \frac{v}{l}\tan u_1,\qquad \dot v = u_2,
    \end{aligned}
\end{equation}
where $\bm x = (x, y, \theta, v)$, $(x,y)\in\mathbb{R}^2$ denotes the location of the vehicle, $v\in\mathbb{R}$ denotes its linear speed, $\theta\in\mathbb{R}$ denotes its heading, $u_1\in\mathbb{R}, u_2\in\mathbb{R}$ are the two controls corresponding to steering wheel angle and acceleration, respectively. $l>0$ denotes the distance between front and rear wheels.

Suppose we have a safety constraint for system (\ref{eqn:bicycle}) defined as:
\begin{equation}\label{eqn:moti_safe}
    (x - x_0)^2 + (y - y_0)^2\geq r^2,
\end{equation}
where $(x_0, y_0)\in\mathbb{R}^2$ denotes the location of the circular obstacle, and $r > 0$ denotes its size.

The relative degree of the safety constraint (\ref{eqn:moti_safe}) is two. Thus, we may use a HOCBF with $m = 2$ as in Def. \ref{def:hocbf} to enforce it. Choosing the class $\mathcal{K}$ functions $\alpha_1, \alpha_2$ as linear functions, the corresponding HOCBF constraint in (\ref{eqn:constraint}) in this case is:
\begin{equation} \label{eqn:moti_hocbf}
    \begin{aligned}
(-2(x-x_0)\sin\theta + 2(y - y_0)\cos\theta)\frac{v^2}{l}\tan u_1 \\+ (2(x - x_0)\cos\theta + 2(y - y_0)\sin\theta)u_2 + 2\dot b(\bm x) + b(\bm x)\geq 0, 
    \end{aligned}
\end{equation}
where $b(\bm x) = (x - x_0)^2 + (y - y_0)^2- r^2$ and $\dot b(\bm x) = 2(x - x_0)v\cos\theta + 2(y - y_0)v\sin\theta$.

Note that the HOCBF constraint (\ref{eqn:moti_hocbf}) is a nonlinear function of $u_1$. Therefore, the eventual CBF-based optimization would be a sequence of NLPs instead of QPs, which makes it hard to solve. One may argue that we can take $\tan u_1$ as a decision variable instead of $u_1$ to make the HOCBF constraint linear in decision variables. However, part of the cost function is to minimize $u_1^2$ instead of $\tan^2 u_1$. As a result, we may have a nonlinear cost function that still makes it become NLPs. Moreover, in some systems, like the quadrotor, the dynamics may include both the control and its quadratic term, which makes the decision variable transformation method further intractable. {We show how we may efficiently guarantee safety for non-affine control systems in this work}.

\subsection{Continuous Optimization Learning Model}

Given a safety constraint $b(\bm x)\geq 0$ whose relative degree is $m$ for non-affine control system (\ref{eqn:system}), we use a HOCBF to enforce it. We still define a sequence of $\psi_i(\bm x), i\in\{1,\dots, m\}$ functions as in (\ref{eqn:functions}), where $\psi_m(\bm x)$ is involved with the control $\bm u$ when combining it with system (\ref{eqn:system}). Therefore, we rewrite $\psi_m(\bm x)$ as $\psi_m(\bm x, \bm u)$, where
\begin{equation} \label{eqn:non_hocbf}
    \psi_m(\bm x, \bm u) = \dot\psi_{m-1}(\bm x, \bm u) + \alpha_m(\psi_{m-1}(\bm x)).
\end{equation}
The above constraint corresponds to the HOCBF constraint in (\ref{eqn:constraint}) for affine-control system (\ref{eqn:affine}). The difference is that the constraint (\ref{eqn:non_hocbf}) is a nonlinear function of $\bm u$, while the HOCBF constraint in (\ref{eqn:constraint}) is a linear function of $\bm u$ given the state $\bm x$.

{In order to address the issue that the constraint (\ref{eqn:non_hocbf}) is a nonlinear function of $\bm u$, we may take an additional derivative of $\psi_m(\bm x, \bm u)$ such that the derivative of $\psi_m(\bm x, \bm u)$ would be linear in $\dot {\bm u}$ following the chain rule.
Formally, we introduce a higher order CBF based on (\ref{eqn:non_hocbf}) in the form}:
\begin{equation}\label{eqn:non_hocbf1}
    \psi_{m+1}(\bm x, \bm u, \dot{\bm u}) = \dot\psi_{m}(\bm x, \bm u, \dot {\bm u}) + \alpha_{m+1}(\psi_{m}(\bm x, \bm u)),
\end{equation}
where $\alpha_{m+1}(\cdot)$ is also a class $\mathcal{K}$ function. The above function is linear in $\dot {\bm u}$. 

In (\ref{eqn:non_hocbf1}), $\dot{\bm u}$ is undefined. We use a neural ODE to model it:
\begin{equation} \label{eqn:node_pi}
    \dot {\bm u} = \pi_{\vartheta}(\bm x, \bm u, \bm I),
\end{equation}
where $\pi_{\vartheta}:\mathbb{R}^{n}\times \mathbb{R}^q\times \mathbb{R}^d \rightarrow\mathbb{R}^q$ is any neural network, such as multi-layer perception (MLP), convolutional neural network (CNN), long-short term memory (LSTM) network, parameterized by $\vartheta$. Recall that $\bm I$ is the observation of system (\ref{eqn:system}).

\begin{figure*}[t]
	\centering
	\includegraphics[scale=1.5]{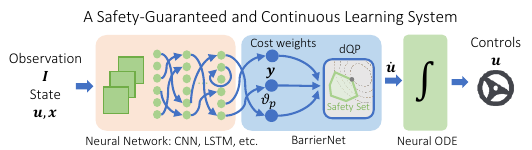}
	\caption{A continuous optimization learning system with safety guarantees. We may define more trainable parameters, such as cost weights, in the BarrierNet (\ref{eqn:bnet}). The model can guarantee safety that is enforced by dCBFs during training or inference in a non-overly-conservative way.  
	}
	\label{fig:model}
\end{figure*}

We can then use training data set to train the neural ODE (\ref{eqn:node_pi}) that makes (\ref{eqn:non_hocbf1}) non-negative. Thus, we can provably show the safety guarantees of this neural ODE controller. However, the neural ODE is not guaranteed to make (\ref{eqn:non_hocbf1}) non-negative due to the uncertainties in training or the model generalization issue. In order to address this, we incorporate (\ref{eqn:non_hocbf1}) into the neural ODE using a BarrierNet, and obtain what we call the continuous optimization learning model:
\begin{equation} \label{eqn:colm}
\begin{aligned}
    \dot {\bm u} &= BarrierNet(\bm x, \bm u, \bm y), \\
    \bm y &= \pi_{\vartheta}(\bm x, \bm u, \bm I),
\end{aligned}
\end{equation}
where $\bm y\in\mathbb{R}^q$ is a latent variable. The above model is continuous in the sense that the number of layers could be interpreted as infinite and we could solve it in continuous time, which is usually compared with discrete learning models (such as CNNs, LSTMs, etc.) that have finite number of layers. For more discussions about continuous versus discrete learning models, please refer to \cite{chen2018neural}. $BarrierNet(\bm x, \bm u, \bm y)$ is defined as
\begin{equation} \label{eqn:bnet}
\begin{aligned}
    BarrierNet &(\bm x, \bm u, \bm y) = \arg \min_{\hat {\bm y}} ||\hat{\bm y} - \bm y||^2,\\
\text{s.t.}&\\
    &\psi_{m+1}^{\vartheta_p}(\bm x, \bm u, \hat{\bm y}) \geq 0,\\
    & b_{min}(\bm u,\hat {\bm y}) \geq 0, \\
    & b_{max}(\bm u,\hat {\bm y}) \geq 0,
\end{aligned}
\end{equation}
where $\hat{\bm y} = \dot {\bm u}$ and $\psi_{m+1}^{\vartheta_p}(\bm x, \bm u, \hat{\bm y}) = \dot\psi_{m}(\bm x, \bm u, \hat{\bm y}) + \vartheta_p\alpha_{m+1}(\psi_{m}(\bm x, \bm u))$ corresponds to the CBF (\ref{eqn:non_hocbf1}) with an additional trainable parameter $\vartheta_p > 0$ that can address the conservativeness of the CBF method \cite{xiao2021bnet}. In addition, we can make $\vartheta_p$ dependent on the observation $\bm I$ such that it is adaptive to the observation. We would also add trainable parameters to all the class $\mathcal{K}$ functions $\alpha_{i}, i\in\{1,\dots,m\}$ in $\psi_{m}(\bm x, \bm u)$. Since the CBF is differentiable in terms of the parameter $\vartheta_p$ in the BarrierNet, we call it a differentiable CBF (dCBF). The model structure is shown in Fig. \ref{fig:model}.

Further, in the BarrierNet (\ref{eqn:bnet}), $b_{min}(\bm u,\hat {\bm y}) = \hat {\bm y} + \bm u - \bm u_{min}$ and $b_{max}(\bm u,\hat {\bm y}) = -\hat {\bm y} + \bm u_{max} - \bm u$ are CBF constraints that enforce the lower control bound $\bm u\geq \bm u_{min}$ and upper control bound $\bm u\leq \bm u_{max}$ as given in (\ref{eqn:control}), respectively. The inequalities are interpreted component-wise. 

Let $odeint(\cdot)$ denote an ODE solver. Now, we provably show the safety guarantees of the continuous optimization learning model (\ref{eqn:colm}).
\begin{theorem}
    Given an initial state $\bm x(0)$ that is safe to system (\ref{eqn:system}) (i.e., $b(\bm x(0))> 0$), and an initial control $\bm u(0)$ such that $\psi_m(\bm x(0), \bm u(0))\geq 0$ as given in (\ref{eqn:non_hocbf}), any control $\bm u(t), t\geq 0$ with
    \begin{equation}
        \bm u(t) = odeint(\text{equation (\ref{eqn:colm})}),
    \end{equation}
 guarantees the safety (i.e., $b(\bm x(t))\geq 0, \forall t\geq 0$) and control bound satisfaction of system (\ref{eqn:system}). 
\end{theorem}
\textbf{Proof:} We first show that the BarrierNet (\ref{eqn:bnet}) is a sequence of QPs given $\bm x$ and $\bm u$. This is important as the ODE solver is intractable during training or inference if it is a sequence of NLPs. The $\psi_m(\bm x, \bm u)$ in (\ref{eqn:non_hocbf}) is a nonlinear function of $\bm u$ for system (\ref{eqn:system}). However, when taking a derivative of $\psi_m(\bm x, \bm u)$, we have that $\dot \psi_m(\bm x, \bm u, \dot{\bm u})$ is a {linear function of $\dot{\bm u}$ following the chain rule}.
Therefore, $\psi_{m+1}^{\vartheta_p}(\bm x, \bm u, \hat{\bm y}) \geq 0$ in (\ref{eqn:bnet}) is a linear function of $\hat{\bm y}$ given $\bm x, \bm u$. The remaining constraints in (\ref{eqn:bnet}) are also linear functions of $\hat{\bm y}$, and the cost is quadratic in $\hat{\bm y}$. Therefore, the BarrierNet (\ref{eqn:bnet}) is a sequence of QPs given $\bm x$ and $\bm u$.

Since $\psi_{m}(\bm x(0), \bm u(0))\geq 0$, $\psi_{m+1}^{\vartheta_p}(\bm x, \bm u, \hat{\bm y}) = \dot\psi_{m}(\bm x, \bm u, \dot {\bm u}) + \vartheta_p\alpha_{m+1}(\psi_{m}(\bm x, \bm u)) \geq 0$ and $\vartheta_p > 0$, following Thm. \ref{thm:hocbf}, we have that $\psi_{m}(\bm x(t), \bm u(t))\geq 0, \forall t\geq 0$. Further, $\psi_{m}(\bm x, \bm u)$ is defined as in (\ref{eqn:non_hocbf}) with $\psi_i(\bm x), i\in\{1,\dots, m-1\}$ recursively defined as in (\ref{eqn:functions}). Since $b(\bm x(0))>0$, we can always find class $\mathcal{K}$ functions $\alpha_i(\cdot), i\in\{1,\dots, m-1\}$ such that $\psi_i(\bm x(0))\geq 0$ \cite{Xiao2019}. Since $\psi_m(\bm x(t), \bm u(t))\geq 0, \forall t\geq 0$ and $\psi_{m-1}(\bm x(0))\geq 0$, we have that $\psi_{m-1}(\bm x(t))\geq 0, \forall t\geq 0$ by Thm. \ref{thm:hocbf}. Recursively, {we can can show that}
$\psi_i(\bm x(t))\geq 0, \forall t\geq 0$ from $i = m-1$ to $i = 0$ using the same argument. Since $b(\bm x) = \psi_{0}(\bm x)$, we have that system (\ref{eqn:system}) is safety guaranteed (i.e., $b(\bm x(t))\geq 0, \forall t\geq 0$).   Along the same line, since the remaining constraints in the BarrierNet (\ref{eqn:bnet}) are CBFs enforcing the control bound (\ref{eqn:control}), we have that the control bound (\ref{eqn:control}) is also satisfied. $\hfill \blacksquare$

The initial control $\bm u(0)$ usually takes a zero vector. The selection of $\bm u(0)$ depends on the generalization of the model.

\begin{remark} [Feasibility guarantees and robustness] 
The BarrierNet may become infeasible at a certain time due to the conflict between the CBFs for safety and control bound. This is still a challenging problem in the CBF method. One possible solution is to find an analytical feasibility constraint that is added to the optimization \cite{Xiao2021}. The system dynamics become affine in $\dot{\bm u}$ inside the neural ODEs. Thus, this feasibility constraint method may still work.  Neural ODEs are generally robust to model uncertainties if the training data is reasonably sampled \cite{chen2018neural}. 
\end{remark}

\textbf{Example revisited.} Now, let's reconsider the example in Sec. \ref{sec:exam}. Within the model (\ref{eqn:colm}), we introduce a higher order of CBF $\psi_2(\bm x,\bm u) := (-2(x-x_0)\sin\theta + 2(y - y_0)\cos\theta)\frac{v^2}{l}\tan u_1 + (2(x - x_0)\cos\theta + 2(y - y_0)\sin\theta)u_2 + 2\dot b(\bm x) + b(\bm x)\geq 0$. The HOCBF constraint corresponding to (\ref{eqn:non_hocbf1}) in this case would take another derivative of $\psi_2(\bm x,\bm u)$, and it is linear in $\dot u_1, \dot u_2$ that we choose as decision variables in (\ref{eqn:bnet}), i.e., $\hat{\bm y} = (\dot u_1, \dot u_2)$. Therefore, the CBF-based optimization becomes a sequence of QPs within the neural ODE (\ref{eqn:colm}).

\subsection{Training of the Model}

In this subsection, we introduce how we may train the continuous optimization learning model (\ref{eqn:colm}). In this work, we focus on the imitation learning method.

In imitation learning, we need to have a nominal controller (such as nonlinear MPC) to generate optimal controls as training labels. The objective is to make the model imitate the nominal controller. A nominal controller for non-affine control system (\ref{eqn:system}) usually involves solving NLPs that are computationally expensive. However, the trained model can be implemented efficiently. The model input includes the system observation $\bm I$, state $\bm x$, and control $\bm u$, as shown in (\ref{eqn:colm}). The inclusion of the control $\bm u$ in the input is to equip the model with short memory. For each of the observation data $\bm I$ with the corresponding state $\bm x$, we can find an optimal control using the nominal controller that label this data. As a result, we can collect training data set in {a diverse set of scenarios}
that system (\ref{eqn:system}) may encounter.

The training of the model (\ref{eqn:colm}) involves solving the following optimization problem
\begin{equation}
(\vartheta^*, \vartheta_p^*) = \arg\min_{\vartheta, \vartheta_p} \mathbb{E}_{\bm x}(\ell(\bm u, \bm u_{nominal}))
\end{equation}
where $\bm u$ is from the solution of (\ref{eqn:colm})---parameterized by $\vartheta, \vartheta_p$ in (\ref{eqn:bnet})---and $\bm u_{nominal}$ is the optimal control from the nominal controller. $\ell(\cdot,\cdot)$ is a similarity measurement, and $\mathbb{E}_{\bm x}$ is the expectation over all the training data. {It has been shown}
that both the neural ODE  and the BarrierNet could be trained using gradient descent in \cite{chen2018neural} and \cite{xiao2021bnet}, respectively, which eventually enables end-to-end (from observation $\bm I$ to control $\bm u$) training of the model.

\textbf{Synthesized Model.} Existing CBF methods assume piece-wise constant controls across discretized time intervals. Our proposed model (\ref{eqn:colm}) could relax this piece-wise constant assumption. This is achieved by incorporating the dynamics (\ref{eqn:system}) into the model (\ref{eqn:colm}):
\begin{equation}
    \begin{aligned}
    \dot {\bm u} &= BarrierNet(\bm x, \bm u, \bm y), \\
    \bm y &= \pi_{\vartheta}(\bm x, \bm u, \bm I), \\
    \dot {\bm x} &= h(\bm x, \bm u),
\end{aligned}
\end{equation}
Then, the neural ODE solver could export safe controls corresponding to each state. The resolution of the control depends on the solver. Specifically, adaptive ODE solvers, such as {\it adaptive\_adams, dopri5} \cite{chen2018neural},
could ensure the solution to be within certain error bounds.

\begin{remark} [Complexity of Training]
    The complexity of the BarrierNet, i.e., the CBF-based QP, is $\mathcal{O}(q^3)$, where $q\in\mathbb{N}$ is the dimension of the decision variable. Therefore, the training complexity of the model (\ref{eqn:colm}) is higher than neural ODEs and other traditional neural networks. The complexity of the BarrierNet can be reduced using the batch QP solving method. Due to the existence of CBFs that guarantee safety during training, the model is harder to train. To simplify the training, We may train the model using two stages: (i) we train the neural ODE without BarrierNet, and (ii) we train the BarrierNet in the model (\ref{eqn:colm}) while fixing all other training parameters. For simple tasks, like the overtaking in driving considered in this paper, we can train the model all at once. 
\end{remark}

\section{CASE STUDIES}

\label{sec:case}

In this section, we consider a high-way driving scenario where the \emph{ego vehicle} overtakes a vehicle in front of it referred to as the \emph{preceding vehicle}. We use {\it fmincon} to solve the NLP for NMPC, use {\it QPFunction} \cite{amos2017optnet} to solve the BarrierNet, and use {\it torchdiffeq} \cite{chen2018neural} with method {\it dopri5} to solve the neural ODE. All the code is implemented in {\it PyTorch}~\cite{paszke2019pytorch} and runs on a {\it AMD Ryzen Threadripper PRO 3975WX 32-Cores} computer.

\noindent\textbf{Problem setup.} The ego vehicle dynamics are non-affine in control, as defined in (\ref{eqn:bicycle}). The ego vehicle has onboard LiDAR observation $\bm I$ (100 distance points around the ego vehicle, as shown in Fig. \ref{fig:setup}(a)) to detect its preceding vehicle.  The preceding vehicle is assumed to have a constant moving speed. The safety constraint between the ego and its preceding vehicle is obtained by a disk-covering approach. In other words, we use an off-the-center disk to cover the preceding vehicle, as shown in Fig. \ref{fig:setup}(a). The disk is designed such that no collisions will happen when the center of the ego vehicle stays outside the disk, and the corresponding safety constraint is $b(\bm x) = (x - x_0)^2 + (y - (y_0 - y_\text{off})) - r^2 \geq 0$, where $(x_0, y_0) \in\mathbb{R}^2$ is the location of the preceding vehicle, and $y_\text{off}\in\mathbb{R}$ is the offset of the disk. The objective of the ego vehicle is to achieve a desired speed while overtaking its preceding vehicle.

\begin{figure}[htbp]
	\vspace{-3mm}
	\centering
	\subfigure[Highway overtaking.]{
		\begin{minipage}[t]{0.24\textwidth}
			\centering
			\includegraphics[width=\textwidth]{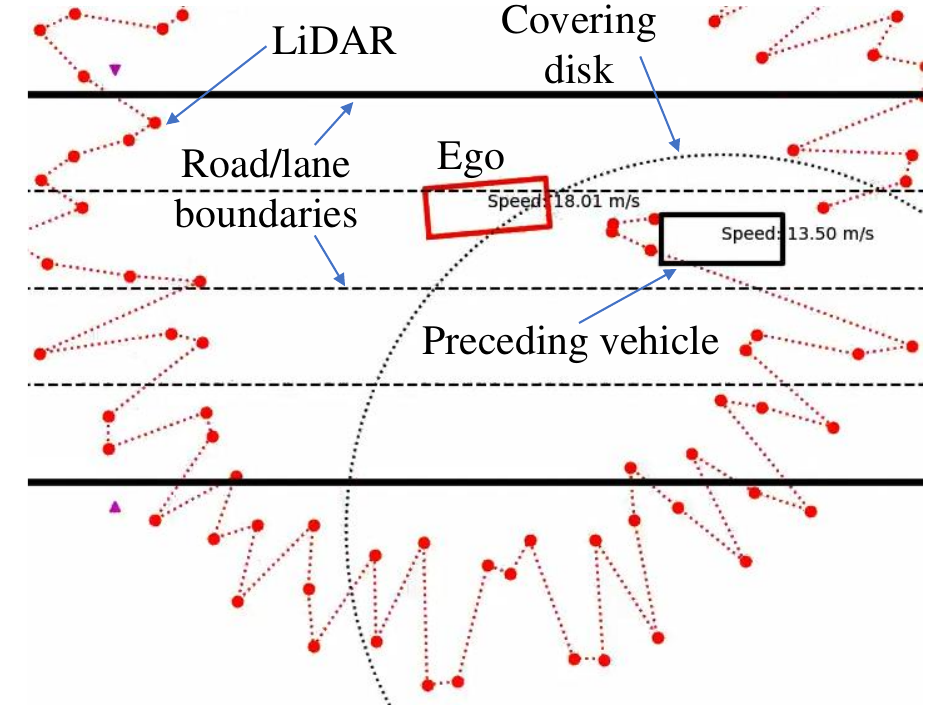}
		\end{minipage}\label{fig:infa}%
	}
	\subfigure[Open-loop validation.]{
		\begin{minipage}[t]{0.22\textwidth}
			\centering
			\includegraphics[width=\textwidth]{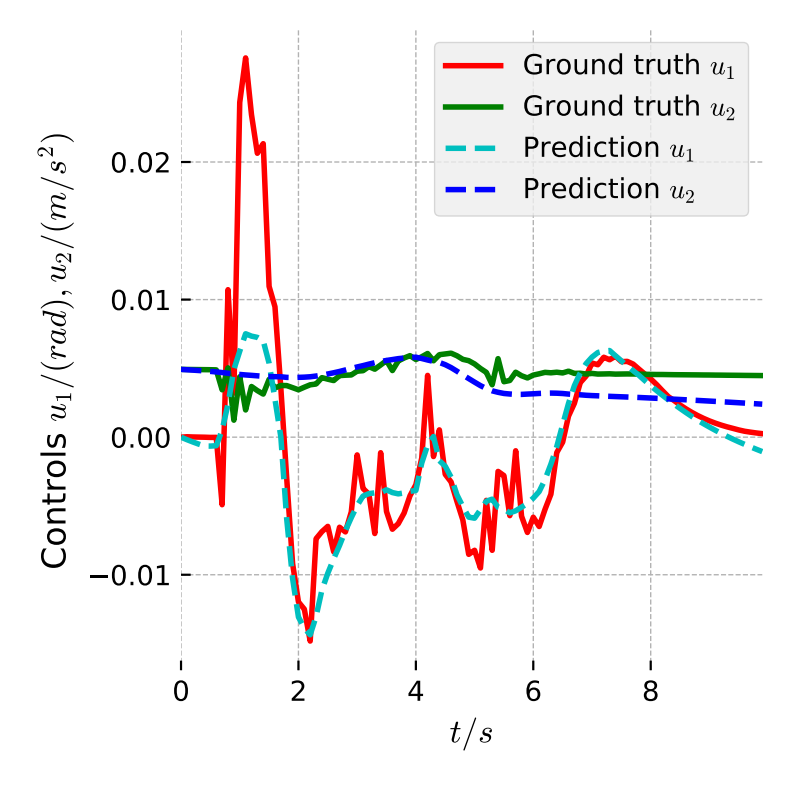}
		\end{minipage}\label{fig:infb}%
	}\hfill		
	\centering
	\caption{Highway overtaking problem setup and end-to-end learning model validation with testing data set.}\label{fig:setup}
\end{figure} 

\noindent\textbf{Training data generation.} We randomly sample initially safe positions with random heading and speed for the ego vehicle around the preceding vehicle. Then, we use a NMPC to find optimal controls for the ego vehicle that can make it safely overtake the preceding vehicle. We collect 201 different initial states, and each initial state will further generate 100 trajectory points using NMPC. The sampling time is 0.1s. Thus, each trajectory corresponds to 10s driving. In summary, the data includes 200 trajectories with optimal control labels as training data set and 1 trajectory as validation data set during training. 

\begin{figure}[htbp]
	\centering
	\subfigure[Closed-loop control profiles.]{
		\begin{minipage}[t]{0.234\textwidth}
			\centering
			\includegraphics[width=\textwidth]{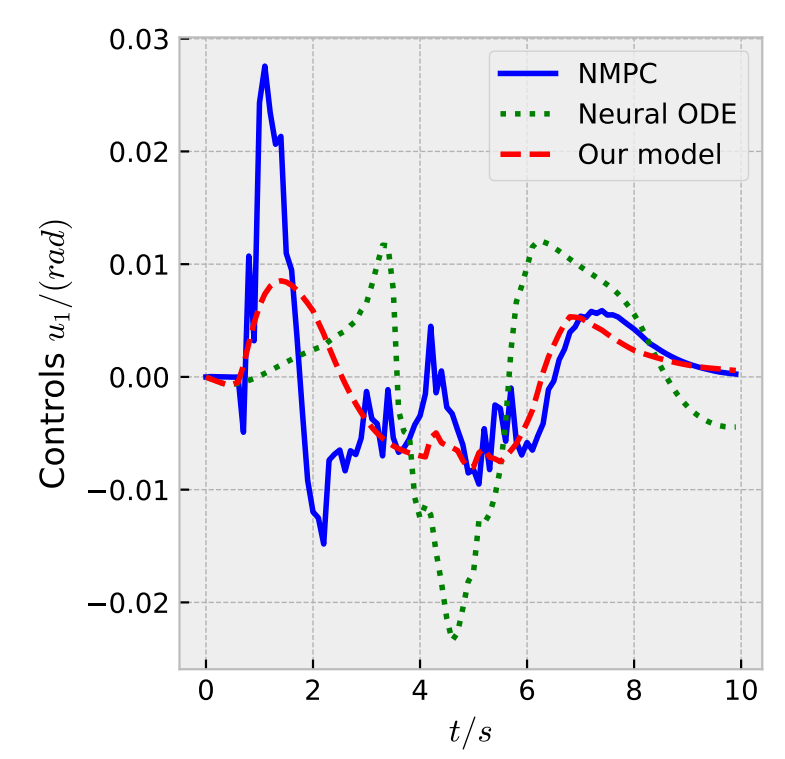}
		\end{minipage}\label{fig:0}%
	}
	\subfigure[Closed-loop safety profiles.]{
		\begin{minipage}[t]{0.23\textwidth}
			\centering
			\includegraphics[width=\textwidth]{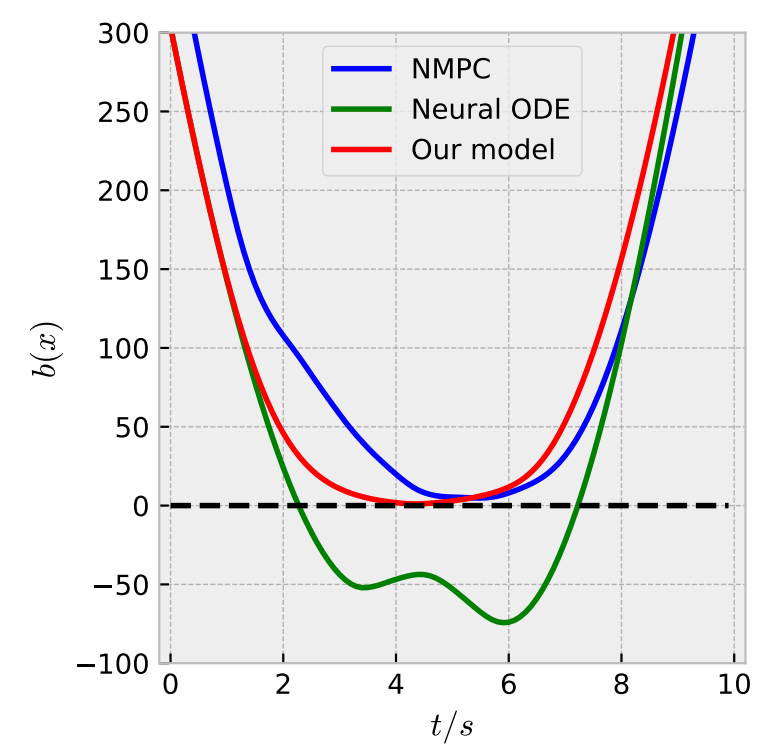}
		\end{minipage}\label{fig:1}%
	}\hfill		
	\centering
	\caption{Highway overtaking closed-loop testing comparisons between NMPC, neural ODE and our models. Safety is guaranteed if $b(\bm x(t))\geq 0, \forall t$}\label{fig:test}
\end{figure} 

\noindent\textbf{Model and training setup.} The model $\pi_{\vartheta}$ in (\ref{eqn:colm}) is defined as a 6-layer MLP with shape (104, 128, 256, 64, 16, 2) followed by a BarrierNet and a neural ODE integration layer, as shown in Fig. \ref{fig:model}. In addition to the LiDAR $\bm I$ and control $\bm u$, we take the heading $\theta$ and speed $v$ of the ego vehicle as input for the model while ignoring its location since this is captured by the LiDAR. We use the {\it RMSprop} in the package {\it torch.optim} as our optimizer during training~\cite{paszke2019pytorch}. The training batch size is 20, and each batch has 10 time sequence trajectory points.

\noindent\textbf{Results.} We first compare the open-loop (Fig. \ref{fig:setup}(b)) and closed-loop (Fig. \ref{fig:test}) testing results. The open-loop testing is based on the test data set without feedback states from dynamics, and it is actually not a good indicator for the performance. In the closed-loop testing, the trajectory from our model can stay close to the NMPC one, while the trajectory from neural ODE could easily violate the safety constraint, as shown in Fig. \ref{fig:test}(b). Both our model and NMPC can guarantee safety.

We further present quantitative comparisons in Table \ref{tab:comp}. The testing results are based on noisy LiDAR (40\% noise magnitude of the LiDAR range), and are from 100 driving overtaking scenarios. Only the neural ODE model fails to guarantee safety, although it is very efficient. The NMPC is very computationally expensive, preventing it from real-time applications. Although linearization is possible in NMPC to decrease its complexity, it may lose guarantees. Our model is computationally efficient under both adaptive ({\it dopri5}) and fixed-time-step ({\it fixed\_adams}) ODE solvers, which is tractable for online safe control. Moreover, our model will not make the system conservative (measured by the average value of the minimum distance with respect to the obstacle among all testing trajectories).  The snapshots of one overtaking example is shown in Fig. \ref{fig:snapshot}, and it shows that collision happens under the neural ODE model, but not in our model.

\begin{table}[ht]
\caption{Self-driving comparisons between NMPC, neural ODE and our model. Items are short for Conservativeness measurement (CONSER.) that is defined by the average value of the minimum distance with respect to the obstacle among all testing trajectories, Safety measurement (SAFETY), Computation time at each iteration under adaptive/fixed-time-step solvers (Computation time), respectively. }
\label{tab:comp}
\begin{center}
\begin{small}
\begin{sc}
\begin{tabular}{p{1.9cm}<{\centering}p{1.5cm}<{\centering}p{1.0cm}<{\centering}p{1.5cm}<{\centering}r}
\toprule
Method &  Conser. ($\geq0$ \& $\downarrow$) & Safety ($\geq 0$)    &Computation time (s) \\
\midrule
NMPC & 4.8 & 4.8  &  1.235 \\
		
Neural ODE   &  $-42.4{\small\pm 4.0}$& -51.6   &  0.007/0.001 \\
		
Ours  &  $1.3{\small\pm 0.2}$ & 0.7  &$0.038$/0.006\\\bottomrule
\end{tabular}
\end{sc}
\end{small}
\end{center}
\end{table}

\begin{figure*}[t]
	\centering
	\includegraphics[scale=0.35]{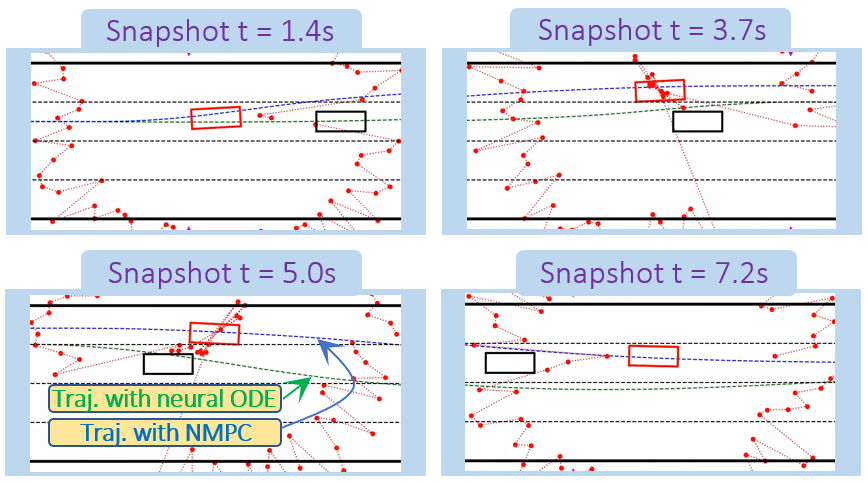}
	\caption{Snapshots of an overtaking simulation (with our proposed model). The trajectory from the neural ODE makes the ego vehicle collide with the preceding vehicle, and the ego vehicle in the snapshots is with safety guarantees as it is controlled by our proposed model. 
	}
	\label{fig:snapshot}
\end{figure*}

\section{CONCLUSION \& FUTURE WORK}

\label{sec:conclusion}

This paper proposes a continuous optimization learning method with safety guarantees for safety-critical systems. The proposed method leverages differentiable control barrier functions and neural ordinary differential equations, and it works efficiently even for non-affine control systems. 
 In the future, we will study the simultaneous modelling of the system dynamics, safety constraints, and control policy using the proposed framework.






\bibliographystyle{IEEEtran}
\bibliography{CBF}

\end{document}